%% file: root.tex
\documentclass[letterpaper, 10 pt, conference]{ieeeconf}  
\usepackage{amsmath}
\usepackage{amsfonts}
\usepackage{graphicx}
\usepackage{color}
\usepackage{tikz}
\usetikzlibrary{positioning}
\usepackage{tikzscale}
\usepackage{import}
\subimport{graphics/layers/}{init}
\usetikzlibrary{backgrounds,calc,shadings,shapes.arrows,arrows,shapes.symbols,shadows,positioning,decorations.markings,backgrounds,arrows.meta}
\usepackage{caption}
\usepackage{stackengine}
\usepackage{hyperref}
\usepackage{nicefrac}
\usepackage{mathtools}

\usepackage{float}
\usepackage{placeins}

\usepackage{array}
\usepackage{booktabs}
\usepackage{multirow}
\usepackage{siunitx}
\usepackage{arydshln}
\makeatletter
\def\adl@drawiv#1#2#3{%
        \hskip.5\tabcolsep
        \xleaders#3{#2.5\@tempdimb #1{1}#2.5\@tempdimb}%
                #2\z@ plus1fil minus1fil\relax
        \hskip.5\tabcolsep}
\newcommand{\cdashlinelr}[1]{%
  \noalign{\vskip\aboverulesep
           \global\let\@dashdrawstore\adl@draw
           \global\let\adl@draw\adl@drawiv}
  \cdashline{#1}
  \noalign{\global\let\adl@draw\@dashdrawstore
           \vskip\belowrulesep}}
\newcommand{\cdashlinelrdotted}[1]{%
  \noalign{\vskip\aboverulesep
           \global\let\@dashdrawstore\adl@draw
           \global\let\adl@draw\adl@drawiv}
  \cdashline{#1}[.4pt/1pt]
  \noalign{\global\let\adl@draw\@dashdrawstore
           \vskip\belowrulesep}}
\makeatother

\definecolor{darkblue}{HTML}{1f4e79}
\definecolor{lightblue}{HTML}{00b0f0}
\definecolor{salmon}{HTML}{ff9c6b}
\definecolor{dodgerblue}{rgb}{0.12, 0.56, 1.0}
\definecolor{frenchblue}{rgb}{0.0, 0.45, 0.73}
\definecolor{green(pigment)}{rgb}{0.0, 0.65, 0.31}
\definecolor{macaroniandcheese}{rgb}{1.0, 0.74, 0.53}
\definecolor{pansypurple}{rgb}{0.47, 0.09, 0.29}
\definecolor{glaucous}{rgb}{0.38, 0.51, 0.71}
\definecolor{hanblue}{rgb}{0.27, 0.42, 0.81}
\definecolor{arylideyellow}{rgb}{0.91, 0.84, 0.42}
\definecolor{newblue}{rgb}{0.56, 0.67, 0.85}
\definecolor{fireenginered}{rgb}{0.81, 0.09, 0.13}
\definecolor{newgreen}{rgb}{0.67, 0.82, 0.57}

\definecolor{newred}{rgb}{0.9, 0.09, 0.1}
\definecolor{newblue}{rgb}{0.56, 0.67, 0.85}
\definecolor{brown}{rgb}{0.6274509804, 0.3215686275, 0.1764705882}
\definecolor{orange}{rgb}{1, 0.9019607843, 0.3921568627}
\definecolor{purple}{rgb}{0.862745098, 0.4705882353, 0.9019607843}

\IEEEoverridecommandlockouts                              

\title{\LARGE \bf
Balancing Spectral, Temporal and Spatial Information \\for EEG-based Alzheimer's Disease Classification}

\author{Stephan Goerttler\authorrefmark{1}\authorrefmark{2}$^{1}$, Fei He\authorrefmark{1} and Min Wu\authorrefmark{2}\\
\authorrefmark{1}Centre for Computational Science and Mathematical Modelling, Coventry University, Coventry, UK\\
\authorrefmark{2}Institute for Infocomm Research, A*STAR, Singapore
\thanks{$^{1}$Stephan Goerttler (goerttlers@uni.coventry.ac.uk) is supported by the A*STAR Research Attachment Program (ARAP).}
}

\begin{document}
\bstctlcite{IEEEexample:BSTcontrol}

\maketitle
\thispagestyle{empty}
\pagestyle{empty}

\begin{abstract}
The prospect of future treatment warrants the development of cost-effective screening for Alzheimer's disease (AD). 
A promising candidate in this regard is electroencephalography (EEG), as it is one of the most economic imaging modalities.
Recent efforts in EEG analysis have shifted towards leveraging spatial information, employing novel frameworks such as graph signal processing or graph neural networks.
Here, we investigate the importance of spatial information relative to spectral or temporal information by varying the proportion of each dimension for AD classification.
To do so, we systematically test various dimension resolution configurations on two routine EEG datasets.
Our findings show that spatial information is more important than temporal information and equally valuable as spectral information.
On the larger second dataset, substituting spectral with spatial information even led to an increase of 1.1\,\% in accuracy, which emphasises the importance of spatial information for EEG-based AD classification.
We argue that our resolution-based feature extraction has the potential to improve AD classification specifically, and multivariate signal classification generally.
\newline
\indent \textit{Clinical relevance}— This study proposes balancing the spectral, temporal and spatial feature resolution to improve EEG-based diagnosis of neurodegenerative diseases.
\end{abstract}

\section{Introduction}
The current clinical gold standard of Alzheimer's disease (AD) diagnosis integrates criteria derived from cognitive testing and positron emission tomography biomarkers \cite{dubois2021clinical}. However, the associated costs prohibit a wider screening of the general population, which would aid in diagnosing the disease at an early stage.
While patients can already benefit from an early diagnosis by knowing about their disease \cite{rasmussen2019alzheimer}, its development is also crucial against the backdrop of prospective future treatments \cite{livingston2019current}.
Electroencephalography (EEG) is an economic and mobile imaging modality touted as a candidate for cost-effective screening of AD \cite{rossini2020early}.

A recent trend has focused on incorporating the graph structure into the data analysis, for example by using graph signal processing \cite{goerttler2023understanding} or graph neural networks \cite{klepl2023graph}. In this work, we explore the relevance of the spatial information for AD classification by balancing it against spectral and temporal information. To this end, we modify the feature resolution along each dimension, while keeping the number of overall features constant.
In particular, the spatial resolution is modified using graph pooling techniques \cite{von2007tutorial}, which allow to preserve graph clusters in the sample. 

In our experiment, we carry out 
the resolution-based feature extraction 
on two routine EEG AD datasets \cite{blackburn2018pilot,miltiadous2023dataset}.
The resulting feature tensors are classified using a support vector machine (SVM). 
This yields the classification performance in dependence on the resolution configuration, which enables us to assess the optimal balance of the dimensions for AD classification.

\section{Methodology}

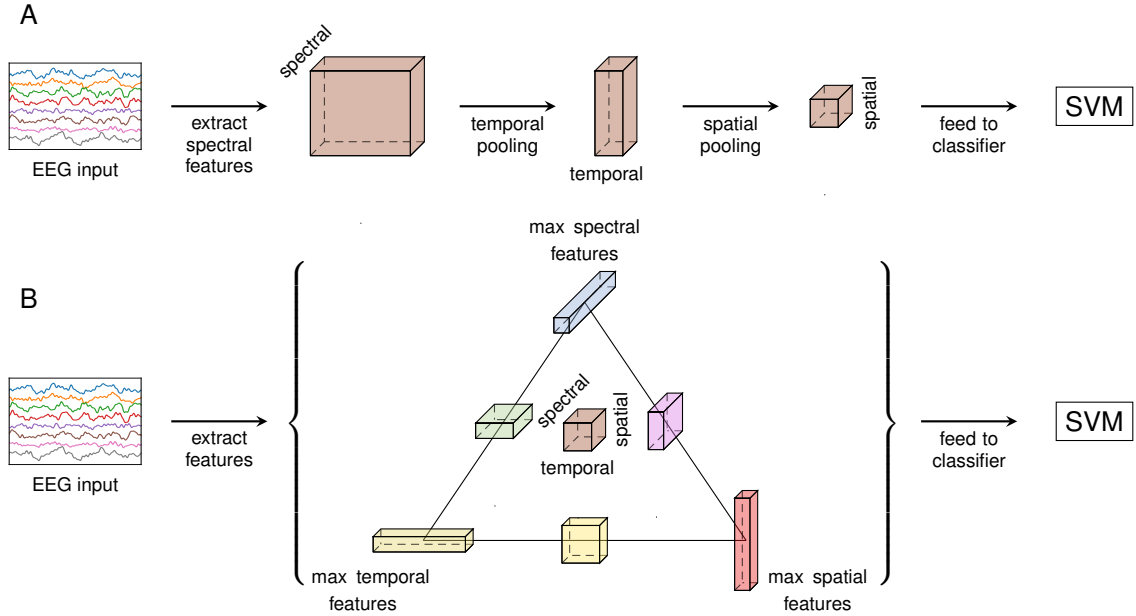
\begin{figure*}[tbh]
    \centering
    \resizebox{0.85\textwidth}{!}{%
    \input{graphics/networks.tikz}%
    }
    \vspace{-0.6cm}
    \caption{Illustration of balancing spectral, temporal and spatial features for classification. (A) Firstly, spectral features are extracted as the spectral density. Note that this step already compresses the temporal dimension. In a second step, the features are pooled along the temporal dimension by averaging. The extent of the pooling determines the temporal resolution. Lastly, the features are pooled spatially using graph pooling, which determines the spatial resolution, before being fed into a support vector machine. (B) When the features are extracted, the resolution along any of the three domains can be varied, changing the shape of the feature matrix as illustrated by the cuboids on the triangle. Importantly, the volume of the displayed cuboids, representing the number of total features, remains constant.}
    \label{fig:method}
\end{figure*}

%

\subsection{Procedure Overview}
The input EEG samples are multivariate signals with a spatial and a temporal dimension.
The goal is to extract feature tensors with three dimensions from these samples, namely the spectral, temporal and spatial dimension, which are subsequently fed into a support vector machine (SVM) classifier. The procedure is illustrated in Figure \ref{fig:method} (A). The first step computes $N_{f,feat}$ power spectral densities (PSDs) for temporally and spatially located time segments. The second and the third step involve pooling the data along the temporal and the spatial dimension, respectively. 
Each transformation can be represented as a computationally inexpensive matrix multiplication.
The extent of the pooling defines the resolution in terms of the number of features for the temporal ($N_{t,feat}$) and the spatial ($N_{g,feat}$) dimension. 
While the number of overall features $N_{feat} = N_{f,feat} N_{t,feat} N_{g,feat}$ is held constant, the number of features for each of the three dimensions can be varied, as illustrated in Figure \ref{fig:method} (B). To allow for many resolution configurations, the number of overall features is set to $N_{feat} = 60=2^2\cdot3\cdot5$ for dataset~I and $N_{feat} = 180=2^2\cdot3^2 
 \cdot5$ for the larger dataset~II.

\subsection{Windowing and Power Spectral Density (PSD) Features}
The extraction of the PSDs builds on Welch's method \cite{welch1967use}.
The input is firstly separated into $N_{seg}$ segments of length $N_{t,seg}$ along the time axis, with each segment overlapping the next by a half. Specifically, the $j$-th segment covers the time interval $\left[j \cdot \lfloor N_{t,seg} / 2\rfloor, j \cdot \lfloor N_{t,seg} / 2\rfloor + N_{t,seg}\right)$.
The length $N_{t,seg}$ of the segments depends both on the target maximum frequency $f_{max}$, the number of frequencies $N_{f,feat}$ as well as the EEG sampling rate $f_s$:
\begin{align}
    N_{t,seg} = \left[N_{f,feat} \frac{f_s}{f_{max}}\right],
\end{align}
where $[\cdot]$ rounds the result to the nearest integer. We set $f_{max}=45\,\mathrm{Hz}$ in our experiments. The number of segments $N_{seg}$ depends on the number of time samples $N_t$ of the input.
Shifting to tensor notation, the two dimensional input matrix $\mathbf{X}$
is separated into $N_{seg}$ segments $\mathbf{X}^{(j)}$ with shape $N_c \times N_{t,seg}$, which we stack horizontally to form a three dimensional tensor $X_{ijk}$ with shape ${N_c\times N_{seg}\times N_{t,seg}}$.

The next steps computes the PSD for each windowed segment. Note that the PSDs are not yet averaged across each window $k$.
This transformation is mathematically given as follows:
\begin{align}
    X_{ijk} \rightarrow \frac{1}{\mathrm{Tr}(W_{kh}^2)} \left| X_{ijk}  W_{kh} S_{hn}^{(\mathrm{DFT})}\right|^2\eqqcolon X_{ijn}^{(f)}\label{eq:psd}.
\end{align}
Here, the matrix $W_{kh}$ is a diagonal matrix with the Hanning window along the diagonal,
\begin{align}
    (W_{kh})_{i,i}=\frac{1}{N_{t,seg}}\sin\left(\frac{i\pi}{N_{t,seg} - 1}\right)^2,
\end{align}
which windows the segments along time.
The matrix $S_{hn}^{(\mathrm{DFT})}$ with shape $N_{t,seg}\times N_{f,feat}$ is the discrete Fourier transform matrix limited to $N_{f,feat}$ columns.

\subsection{Temporal Pooling}
The tensor $X_{ijn}^{(f)}$, computed in \eqref{eq:psd}, is subsequently pooled along the temporal dimension. The number of pooling groups $N_{t,feat}$, with $1\leq N_{t,feat} \leq N_{t,seg}$, defines the number of temporal features and thereby the temporal resolution. 
To form the groups with equal length, we can assign a group index $n_t(j)$ to each temporally located segment $j$ using the floor function $\lfloor\cdot\rfloor$:
\begin{align}
    n_t(j) = \left\lfloor \frac{j}{\lfloor N_{t,seg} / N_{t,feat}\rfloor} \right\rfloor.
\end{align}
This allows us to define a group assignment matrix $S_{jm}^{(\mathrm{time})}$ as follows:
\begin{align}
    \left(S_{jm}^{(\mathrm{time})}\right)_{i,j} &=     
    \begin{cases}
      \nicefrac{1}{N_{t,grp}(j)} & \text{if $j=n(i)$}\\
      0 & \text{else}
    \end{cases} \label{eq:Stime} 
    \\
    N_{t,grp}(j) &= \#\{n_t(k)=j,\forall k\},
\end{align}
where $N_{t,grp}(j)$ counts the number of elements in the respective group. 
Multiplication with the group assignment matrix essentially averages the windows belonging to each of the groups across time. 
The tensor with the extracted spectral and temporal features is then computed using the following transformation:
\begin{align}
    X_{ijn}^{(f)} \rightarrow X_{ijn}^{(f)} S_{jm}^{(\mathrm{time})} \eqqcolon X_{imn}^{(ft)}.
\end{align}

\subsection{Spatial Pooling}
We employ graph spectral clustering \cite{von2007tutorial}, or graph pooling, to pool the feature tensor $X_{imn}^{(ft)}$ along the spatial dimension. The goal of the graph pooling is to control the number of spatial features $N_{g,feat}$, which define the spatial resolution. 
%
To do so, we firstly retrieve the overall graph structure of each training set data-driven using the functional connectivity. Specifically, we compute the graph's adjacency matrix $\mathbf{A}$ pairwise as the Pearson correlation between channels, meaning that $\mathbf{A}$ is symmetric and zero on its diagonal.
Secondly, we compute the Laplacian matrix $\mathbf{L}=\mathbf{D} - \mathbf{A}$ from the adjacency matrix, where $\mathbf{D}=\mathrm{diag}(\mathbf{A}\cdot \mathbf{1})$ denotes the degree matrix.
Thirdly, we compute the first $N_{c,feat}$ eigenvectors of $\mathbf{L}$, stack the eigenvectors horizontally to form the matrix $\mathbf{U}$ and read out the $N_c$ rows as vectors $y_i$ of length $N_{c,feat}$. The vectors $y_i$ define points which are clustered using the k-means algorithm. This ultimately yields a cluster, or group, index $n_g(i)$ for each channel $i$. In analogy to equation \eqref{eq:Stime}, we can define a group assignment matrix $S_{il}^{(\mathrm{graph})}$ using the index mapping $n_g$:
\begin{align}
    \left(S_{il}^{(\mathrm{graph})}\right)_{i,j} &=     
    \begin{cases}
      \nicefrac{1}{N_{g,grp}(j)} & \text{if $j=n_g(i)$}\\
      0 & \text{else}
    \end{cases} \label{eq:Sgraph} 
    \\
    N_{g,grp}(j) &= \#\{n_g(k)=j,\forall k\},
\end{align}
yielding the final transformation:
\begin{align}
    X_{imn}^{(ft)} \rightarrow X_{imn}^{(ft)} S_{il}^{(\mathrm{graph})} \eqqcolon X_{lmn}^{(ftg)}.
\end{align}
The resulting feature tensor $X_{lmn}^{(ftg)}$ has shape $N_{g,feat}\times N_{t,feat}\times N_{f,feat}$ and is flattened before being fed into the SVM classifier.








\subsection{Support Vector Machine (SVM) Classification}
We used a non-linear SVM with commonly used parameters to classify the extracted features. Specifically, the radial basis function kernel coefficient is given by $\gamma = 1 / (N_f \,\mathrm{Var}(X))$, where $X$ is the feature vector, and the regularisation strength by $C=1$. This study focuses on comparing feature extraction parameters, which is why we did not perform hyperparameter optimisation.
To test our model, we used 10-fold cross validation. The multiple samples of each patient were strictly kept in one fold for both datasets to avoid data leakage.

\section{Experiments}

\subsection{Dataset I}
The first EEG dataset used in this study was acquired from 20 Alzheimer's disease patients and 20 healthy controls at a sampling rate of 2048$\,$Hz with a modified 10-20 placement method \cite{blackburn2018pilot}. Patients were instructed to close their eyes during the measurement. For every patient, two or three sections of 12 seconds were selected from the recording by a clinician, resulting in 119 samples. To cancel out volume conduction artifacts, a bipolar montage was used, resulting in 23 channels. The full description of the dataset can be found in Blackburn et al. \cite{blackburn2018pilot}. 
We further downsampled the samples by a factor of 4 to a sampling rate of $512\,\mathrm{Hz}$ to match the sampling rate in dataset II. The final shape of each sample is $N_c \times N_t=23\times 6,145$.

\subsection{Dataset II}
The second, publicly available\footnote{\href{https://openneuro.org/datasets/ds004504/versions/1.0.6} {https://openneuro.org/datasets/ds004504/versions/1.0.6}} EEG dataset comprises 36 Alzheimer's disease patients and 29 healthy controls \cite{miltiadous2023dataset}.
It was recorded at a sampling rate of $500\,\mathrm{Hz}$ with 19 scalp electrodes placed using the 10-20 system and two reference electrodes. As in dataset I, patients were instructed to keep their eyes closed during the 13-14 minute measurements. The EEG samples were filtered and re-referenced using the reference electrodes, and artefact detection methods were employed. More details about the dataset and the preprocessing are described in Miltiadous et al. \cite{miltiadous2023dataset}.
We partitioned the preprocessed EEG samples into 60 seconds-long samples, excluding samples that contained artefacts. This resulted in overall 656 EEG samples with shape $N_c \times N_t=19\times 30,000$.

\subsection{Results}
\begin{figure*}[!t]
    \centering
    \resizebox{1.0\textwidth}{!}{%
    \begin{tikzpicture}[every node/.style={inner sep=0,outer sep=0}]
    \sffamily
      \node[anchor=west] (image) at (0,0) {\includegraphics[width=0.46\textwidth,trim={0.95cm 0 1.67cm 0},clip]{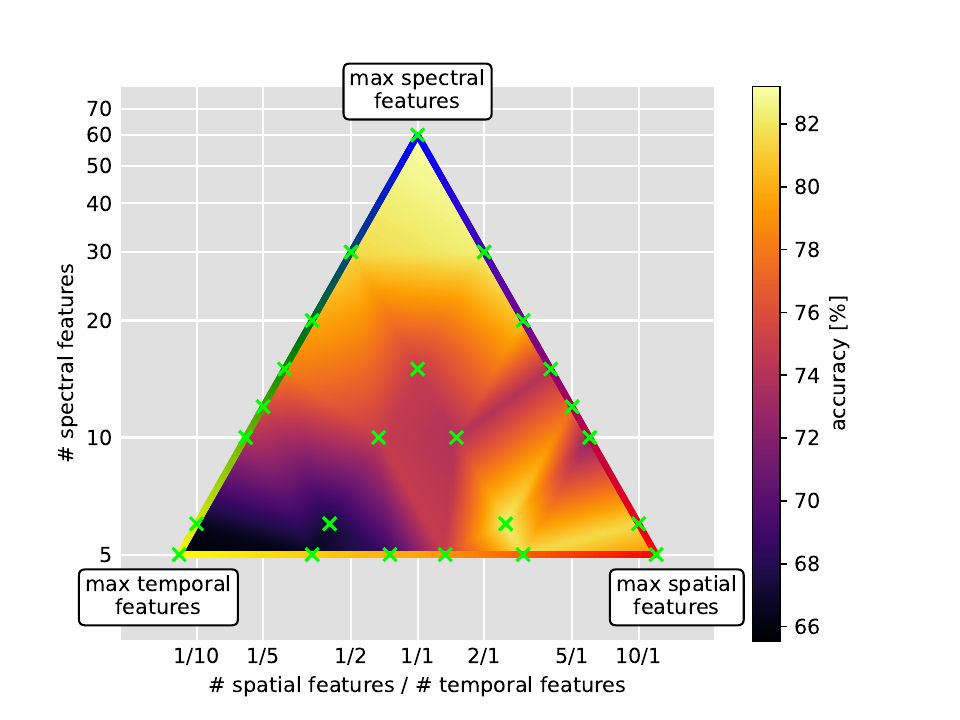}};
        \node[anchor=west] (image) at (9,0) {\includegraphics[width=0.46\textwidth,trim={0.75cm 0 1.87cm 0},clip]{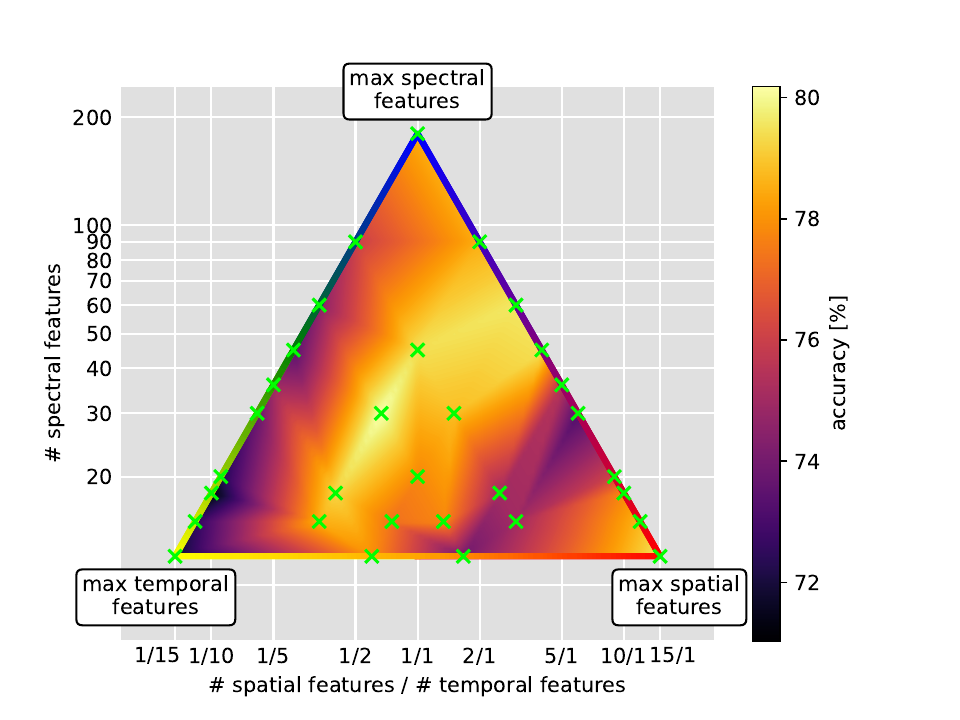}};
        \node (caption) at (0.4,3.2){A};
        \node (caption) at (9.525,3.2){B}; 
    \end{tikzpicture}
    }
    \caption{Linearly interpolated accuracy in dependence on the feature resolution configuration for dataset I (A) and II (B). The green crosses mark the experimentally tested configurations. The accuracy along the triangle edges is separately plotted in Figure \ref{fig:results_edge}. 
    Both datasets reveal similar levels of accuracy for maximal spectral and graph features, as well as poor accuracy for maximal time features. They further both reveal an accuracy level between the maximal spectral and graph features. 
    In the centre of the available configuration space, corresponding to more balanced resolution configurations, the performance diverges across the two datasets.
    }
    \label{fig:results_triangle}
\end{figure*}

Figure \ref{fig:results_triangle} shows the accuracy in dependence on the feature configuration. While there are three variable parameters, namely the number of features along each dimension, the constraint that the overall number of features is constant reduces the number of free parameters to two. We selected the ratio of the number of graph features to the number of time features as one variable, and the number of spectral features as the second, while the accuracy is depicted in colour as an interpolated contour plot. The results show that maximising the number of time features yields the lowest accuracy. Conversely, maximising spatial features yields comparable accuracy to maximising spectral features, and even an increase of 1.1\,\% on the larger dataset II.
Interestingly, more balanced configurations result in markedly different performances across the two datasets.

\begin{figure}[!t]
  \centering
  {\includegraphics[width=0.99\columnwidth]{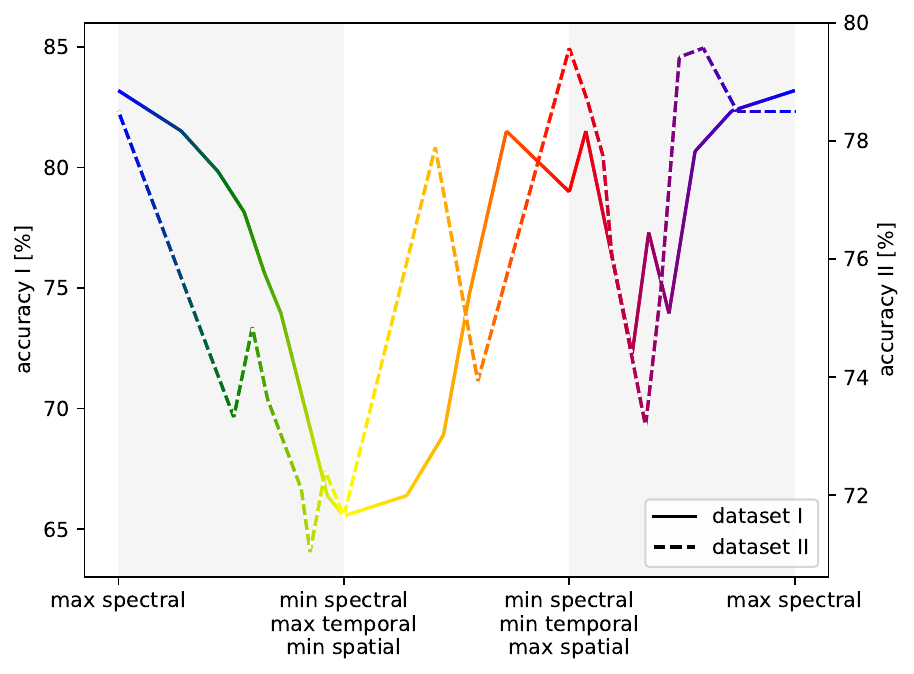}}
  \caption{Accuracy along the triangle edges depicted in Figure \ref{fig:results_triangle}. The colour of the curve allows to retrieve the feature resolution configuration from Figure \ref{fig:results_triangle}. The curve has a large accuracy valley at maximal temporal information (yellow section), but also a smaller accuracy valley between the maximal spatial and spectral configuration (magenta section).}
  \label{fig:results_edge}
\end{figure}

Figure \ref{fig:results_edge} shows the accuracy along the edge of the triangular configuration space, which includes all configurations for which at least one of the three variables is minimal. The accuracy curves of the two datasets are strikingly similar. They both depict a large valley for a maximal number of time features, and a small valley for configurations in between maximal graph and maximal spectral features.

\section{Discussion}
We observed that high temporal feature resolution at the expense of spectral or spatial resolution
led to poorer performance on two EEG-based AD classification tasks, possibly because the dynamics are not synchronised in the experiments. 
On the other hand, favouring spatial over spectral resolution led to comparable or even higher performances, suggesting that spatial information can offset the loss of spectral information typically crucial for EEG.
However, the accuracy drop for in-between configurations, visible in both datasets, indicates that balancing spectral and spatial features is not trivial,  but rather relies on a suitable combination of frequency spacing and graph partition. 
The performance of configurations that balance all three dimensions diverges across the two datasets, which suggests that these configurations are more susceptible to differences in the EEG setup.

%
%
%
%
In conclusion, this study introduced a feature extraction tool that enables the arbitrary balancing of spatial, spectral, and temporal information in the extracted features.
Our results highlight the potential benefits of leveraging spatial information for multivariate signal classification. We argue that while temporal information appeared redundant in our experiment, it may similarly prove valuable for classification tasks involving time-locked signals.

\bibliographystyle{IEEEtran}
\bibliography{refs}

\end{document}

%% file: graphics/layers/init.tex
\usetikzlibrary{quotes,arrows.meta}
\usetikzlibrary{positioning}

\usepackage{Box}

%% file: graphics/networks.tikz
\begin{tikzpicture}
    \sffamily

    \def\offsetexample{4.1}
    \def\depthfactor{1.3}
    \def\cmfactor{4.6}
    \def\arrowlength{1.25}
    \def\factorshift{0.2}
    \def\triangleright{2.1}
    \def\triangleup{1.55}
    \def\centershift{2.5}
    \def\trianglestart{4.125}
    \def\textfactor{1.42}


    \node[label=below:\scriptsize EEG input,inner sep=0pt] (EEG_1) at (0,\offsetexample)
    {\includegraphics[width=1.8cm, height=1.2cm, angle=0]{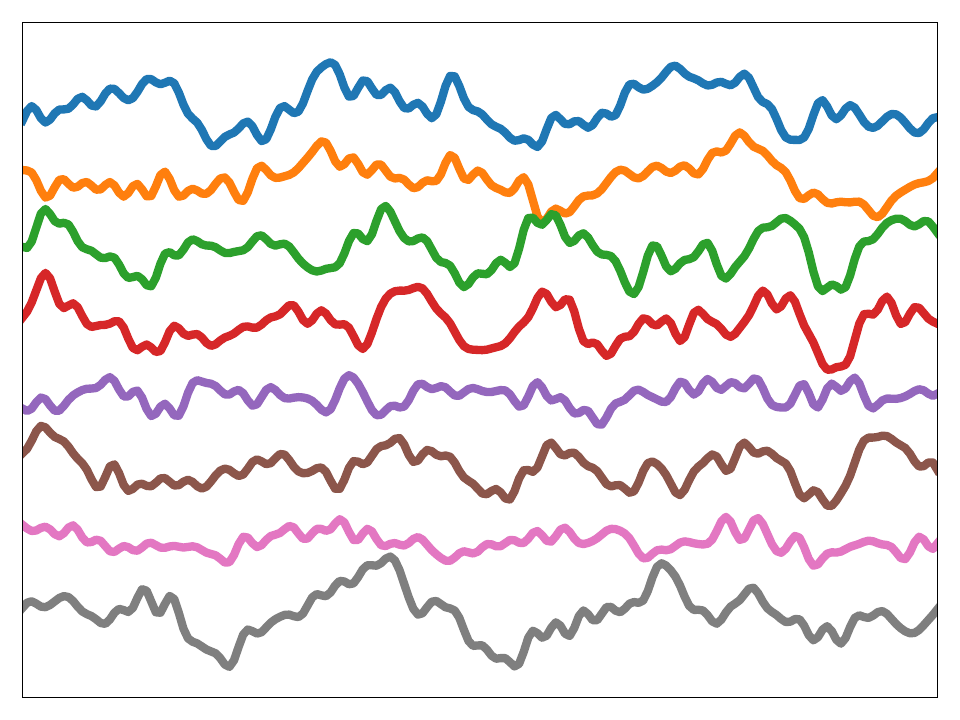}};
    
    \draw[-stealth,thick] (1.25, \offsetexample) -- (1.25 + \arrowlength,\offsetexample) node[text width=1.2cm,align=center, font={\scriptsize}, midway,below] {extract spectral features};
    
    \pic[shift={(0.25,0)}] at (2.9, \offsetexample) {Box={name=conv1,fill=brown,height=1.2*\cmfactor,width=6.5,depth=1.81712059283 * \depthfactor}};
    \node[text width=1.5cm, align=center, rotate=45] at (3, 0.7 + \offsetexample) {\scriptsize spectral};
    
    \draw[-stealth,thick] (5, \offsetexample) -- (5 + \arrowlength,\offsetexample) node[text width=1.2cm,align=center, font={\scriptsize}, midway,below] {temporal pooling};
    
    \pic[shift={(0.25,0)}] at (6.6, \offsetexample) {Box={name=conv1,fill=brown,height=1.2*\cmfactor,width=1.81712059283,depth=1.81712059283 * \depthfactor}};
    \node[text width=1.5cm, align=center] at (6.6 + 0.25, -0.9 + \offsetexample) {\scriptsize \,\,\,temporal};
    
    \draw[-stealth,thick] (7.9, \offsetexample) -- (7.9 + \arrowlength,\offsetexample) node[text width=1.2cm,align=center, font={\scriptsize}, midway,below] {spatial pooling};
    
    \pic[shift={(0.25,0)}] at (9.4, \offsetexample) {Box={name=conv1,fill=brown,height=1.81712059283,width=1.81712059283,depth=1.81712059283 * \depthfactor}};
    \node[text width=1.5cm, align=center, rotate=90] at (9.6 + 0.25 + 0.5, \offsetexample) {\scriptsize \,\,spatial};

    \draw[-stealth,thick] (11, \offsetexample) -- (12.25, \offsetexample) node[text width=1.2cm,align=center, font={\scriptsize}, midway,below] {feed to classifier};

    \node[draw,align=left,anchor=west] at (12.75,\offsetexample) {SVM};

    \node[label=below:\scriptsize EEG input,inner sep=0pt] (EEG_1) at (0,0)
    {\includegraphics[width=1.8cm, height=1.2cm, angle=0]{figures/Alzheimers_EEG.pdf}};

    \draw[-stealth,thick] (1.25, 0) -- (2.5,0) node[text width=1.2cm,align=center, font={\scriptsize}, midway,below] {extract features};
    \draw[-stealth,thick] (11, 0) -- (12.25, 0) node[text width=1.2cm,align=center, font={\scriptsize}, midway,below] {feed to classifier};

    \node[draw,align=left,anchor=west] at (12.75,0) {SVM};
    
    
    \node (left-paren) [right = 2.6cm] {$\left\{\vphantom{\includegraphics[width=0pt, height=2.25cm]{figures/Alzheimers_EEG.pdf}}\right.$};
    \node (right-paren) [right = 10.25cm] {$\left.\vphantom{\includegraphics[width=0pt, height=2.25cm]{figures/Alzheimers_EEG.pdf}}\right\}$};

    \draw[-,thin] (-\triangleright+\trianglestart+\centershift, -\triangleup) -- (\trianglestart+\centershift, \triangleup);
    \draw[-,thin] (\triangleright+\trianglestart+\centershift, -\triangleup) -- (\trianglestart+\centershift, \triangleup);
    \draw[-,thin] (-\triangleright+\trianglestart+\centershift, -\triangleup) -- (\triangleright+\trianglestart+\centershift, -\triangleup);

    \node[text width=1.6cm, align=center] at (-\triangleright * \textfactor+\trianglestart+\centershift + 0.2, -\triangleup * \textfactor + 0.16) {\scriptsize max temporal};
    \node[text width=1.5cm, align=center] at (-\triangleright * \textfactor+\trianglestart+\centershift + 0.1, -\triangleup * \textfactor - 0.16) {\scriptsize features};
    \node[rotate=0, text width=1.5cm, align=center] at (\triangleright * \textfactor+\trianglestart+\centershift + 0.05, -\triangleup * \textfactor + 0.16) {\scriptsize max spatial};
    \node[rotate=0, text width=1.5cm, align=center] at (\triangleright * \textfactor+\trianglestart+\centershift + 0.05, -\triangleup * \textfactor - 0.16) {\scriptsize features};
    \node[rotate=0, text width=1.5cm, align=center] at (\trianglestart+\centershift, \triangleup * \textfactor + 0.2 + 0.16 - 0.05) {\scriptsize max spectral};
    \node[rotate=0, text width=1.5cm, align=center] at (\trianglestart+\centershift, \triangleup * \textfactor + 0.2 - 0.16 - 0.05) {\scriptsize features};

    \pic[shift={(\trianglestart+\centershift-0.5*\factorshift,\triangleup,0)}] at (0, 0) {Box={name=conv2,fill=newblue,height=1,width=1,depth=6*\depthfactor}};
    
    \pic[shift={(\trianglestart+\centershift-1.81712059283/2*\factorshift,-0.07179676972449078 * \triangleup,0)}] at (0, 0) {Box={name=conv1,xlabel={"temporal","dummy"},fill=brown,height=1.81712059283,width=1.81712059283,depth=1.81712059283 * \depthfactor}};
    
    \node[text width=1.5cm, align=center, rotate=45] at (\trianglestart+\centershift - 0.27, 0.255) {\scriptsize spectral};
    
    \node[text width=1.5cm, align=center, rotate=90] at (\trianglestart+\centershift + 0.49, 0.025) {\scriptsize spatial};

    \pic[shift={(-\triangleright/2+\trianglestart+\centershift-2.44948974278/2*\factorshift,0,0)}] at (0,0) {Box={name=conv3,fill=newgreen,height=1,width=2.44948974278,depth=2.44948974278 * \depthfactor}};
    
    \pic[shift={(-\triangleright+\trianglestart+\centershift-6/2*\factorshift,-\triangleup,0)}] at (0,0) {Box={name=globalavg,fill=arylideyellow,height=1,width=6,depth=1 * \depthfactor}};
    
    \pic[shift={(+\triangleright+\trianglestart+\centershift-1/2*\factorshift,-\triangleup,0)}] at (0,0) {Box={name=out_1,fill=newred,height=6,width=1,depth=1 * \depthfactor}};

    \pic[shift={(\trianglestart+\centershift-2.44948974278/2*\factorshift,-\triangleup,0)}] at (0,0) {Box={name=conv3,fill=orange,height=2.44948974278,width=2.44948974278,depth=1 * \depthfactor}};
    
    \pic[shift={(\trianglestart+\centershift+\triangleright/2-1/2*\factorshift,0,0)}] at (0,0) {Box={name=conv3,fill=purple,height=2.44948974278,width=1,depth=2.44948974278 * \depthfactor}};





    \node (caption) at (-0.6,1.2+\offsetexample){A};
    \node (caption) at (-0.6,1.6){B};
    
\end{tikzpicture}